\newcommand{\CLASSINPUTtoptextmargin}{0.75in}
\newcommand{\CLASSINPUTbottomtextmargin}{1in}
\documentclass[conference,letterpaper]{IEEEtran}
\IEEEoverridecommandlockouts

\usepackage{amsmath,amssymb,amsfonts}
\usepackage{algorithmic}
\usepackage{graphicx}
\graphicspath{{./Figures/}}
\usepackage{textcomp}
\usepackage{xcolor}
\usepackage{subfiles} 
\usepackage{subcaption}
\usepackage{fancyhdr}

\usepackage{color,soul}
\usepackage{cite}
\usepackage{amsmath}
\usepackage{multirow}

\def\BibTeX{{\rm B\kern-.05em{\sc i\kern-.025em b}\kern-.08em
    T\kern-.1667em\lower.7ex\hbox{E}\kern-.125emX}}
\fancypagestyle{firststyle}{
	\fancyhf{}
	\fancyhead[L]{ O. Kanhere, K. F. Nieman, and S. S. Ghassemzadeh, ``Indoor-Office Large-Scale Wireless Channel Characterization in cmWave/FR3 Spectrum,'' \textit{in GLOBECOM 2025 - 2025 IEEE Global Communications Conference, Taipei, Taiwan, Dec 2025, pp.
1–6. } }    %

}
\begin{document}
\bstctlcite{IEEEexample:BSTcontrol}

\title{Indoor-Office Large-Scale Wireless Channel Characterization in cmWave/FR3 Spectrum

}
\author{\IEEEauthorblockN{Ojas Kanhere, Karl. F. Nieman, and
Saeed S. Ghassemzadeh}
\IEEEauthorblockA{AT\&T Labs, Austin, TX USA\\
Email: \{ok067s, kn2644, sg2121\}@att.com
}}
\maketitle

\thispagestyle{firststyle}

\begin{abstract}
This paper presents comprehensive findings on the characterization of Indoor Hotspot channel parameters, derived from an extensive experimental campaign conducted at 6.9, 8.3, and 14.5 GHz in a commercial office building. Extensive measurements were carried out in diverse indoor office settings, including cubicles, conference rooms, hallways, and laboratory spaces across four floors. The path loss, shadow fading, delay spread, and angular spread was modeled. Our results offer significant insights into the attenuation and dispersion characteristics of wireless signals in diverse indoor settings in the centimeter-wave frequency band, and can be used for improving indoor network design and performance in commercial buildings.

\end{abstract}


\section{Introduction}
\label{sec:intro}
The rapid advancement of wireless communication technologies has brought about significant changes in our daily lives, with high-speed, reliable connectivity now essential. To meet the growing demand for spectrum required to achieve such high data rates with good coverage, the newly introduced upper mid-band Frequency Range 3 (FR3) in the centimeter-wave spectrum (7-24 GHz) was introduced. The Indoor Hotspot (InH) environment, a possible deployment scenario for FR3 spectrum, is characterized by a complex interplay of various physical factors, such as walls, partitions, furniture, and the presence of electronic devices and human bodies. These elements can cause significant signal attenuation, reflection, diffraction, and scattering, which impact the overall performance of wireless communication systems. Therefore, prior to deployment, accurately modeling and characterizing the large-scale wireless channel parameters in such environments is crucial for the effective design and deployment of next-generation wireless networks. 

There have been numerous indoor FR3 measurements conducted that have significantly contributed to our understanding of InH wireless channel behavior \cite{Bohdanowicz_1999, Janssen_1996,Batalha_2019,Kim_2014,Shakya_2024_01,Shakya_2025}. In \cite{Bohdanowicz_1999}, the authors studied indoor line-of-sight (LOS) and obstructed LOS propagation characteristics at 17 GHz, while in \cite{Janssen_1996}, the authors measured the indoor wireless channel at 2.4, 4.75, and 11.5 GHz. However the angular characteristics of the channel were not measured in either study. In \cite{Batalha_2019}, the authors conducted indoor corridor and office measurements at 8, 9, 10, and 11 GHz, however non-line-of-sight (NLOS) locations were not evaluated and neither angular spread nor delay spread was characterized. In \cite{Kim_2014}, authors conducted indoor wideband measurements at 11 GHz over 220 snapshots captured every 14 cm. In \cite{Shakya_2024_01, Shakya_2025}, authors conducted indoor measurements at 6.75 and 16.95 GHz, however since a total of 20 locations were measured, more insight could be gained by conducting more extensive indoor measurement campaigns. 

This paper attempts addressing the gap in  understanding of the indoor channel parameters and their frequency variation over the FR3 band. The channel is measured at same locations across three FR3 frequencies, namely, 6.9, 8.3, and 14.5 GHz. The channel path loss, shadow fading, angular spread, and delay spread was measured at 653 unique measurement locations over four floors of a commercial office building.

The remainder of this paper is organized as follows. In Section \ref{sec:HW_Experiments} we describe the measurement equipment, procedures and data processing needed to extract parameters from measured data. Section \ref{sec:KeyFindings} covers our key findings of the channel parameters followed by conclusions in Section \ref{sec:conclusions}.

\section{Measurement Campaign}
\label{sec:HW_Experiments}
In this section, we shall go over the channel sounder configuration, indoor environment, experimental procedure and data processing required for conducting our indoor FR3 channel measurement campaign.

\subsection{The Channel Sounder} 
The channel sounder used for this work \cite{Nieman_2025_01} was designed to accurately sample the wireless channel within the FR3 frequency range. A Zadoff Chu (ZC) sounding sequence was used, modulated using orthogonal frequency domain modulation (OFDM) with a cyclic prefix. The ZC sequence was repeated four times in the time domain to yield an averaging gain of 6 dB. The ZC sequence, produced by software-defined radios (SDRs), was upconverted to an intermediate frequency (IF). The IF signal was then upconverted to RF, amplified using an RF power amplifier, and transmitted via a 10 dBi standard gain horn antenna. Note that each frequency band (6.9 GHz, 8.3 GHz, and 14.5 GHz) had a separate RF upconverter, which produced an RF signal with a maximum effective isotropic radiated power (EIRP) of +43 dBm in each frequency band. Power was reduced using a variable attenuator to avoid saturating the receiver (RX) at short distances. 
\begin{figure}[t]
    \centering
    \subfloat[Office Layout]{
        \includegraphics[trim={0.5cm 0 0 0},width=0.98\columnwidth]{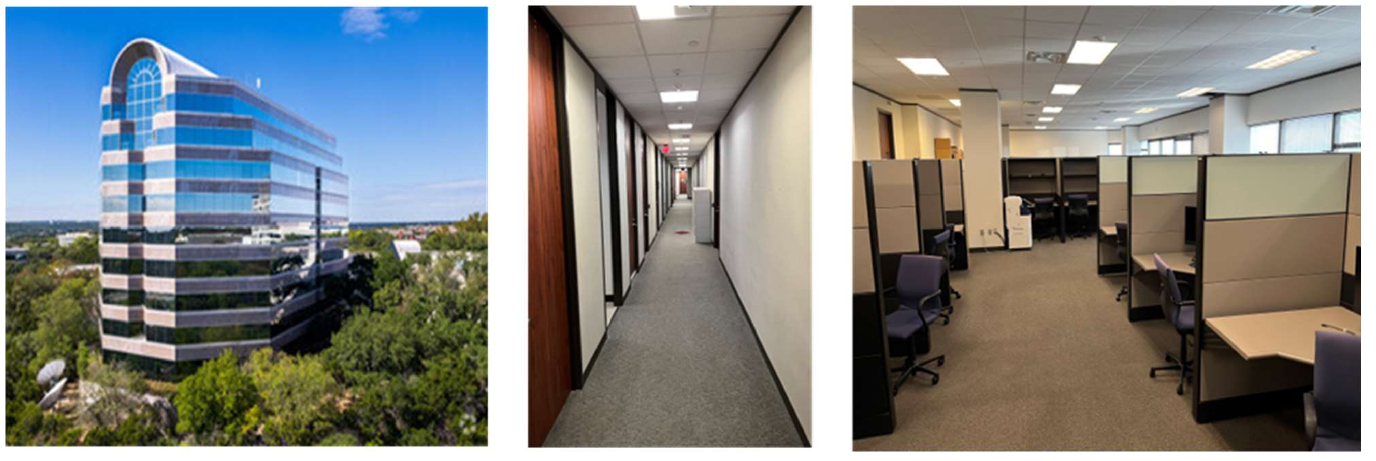} 
        \label{fig:Arbor_bldg}
    } \\
    
    \subfloat[Typical Floor Plan]{
        \includegraphics[trim={0.5cm 0 0 0},width=0.98\columnwidth]{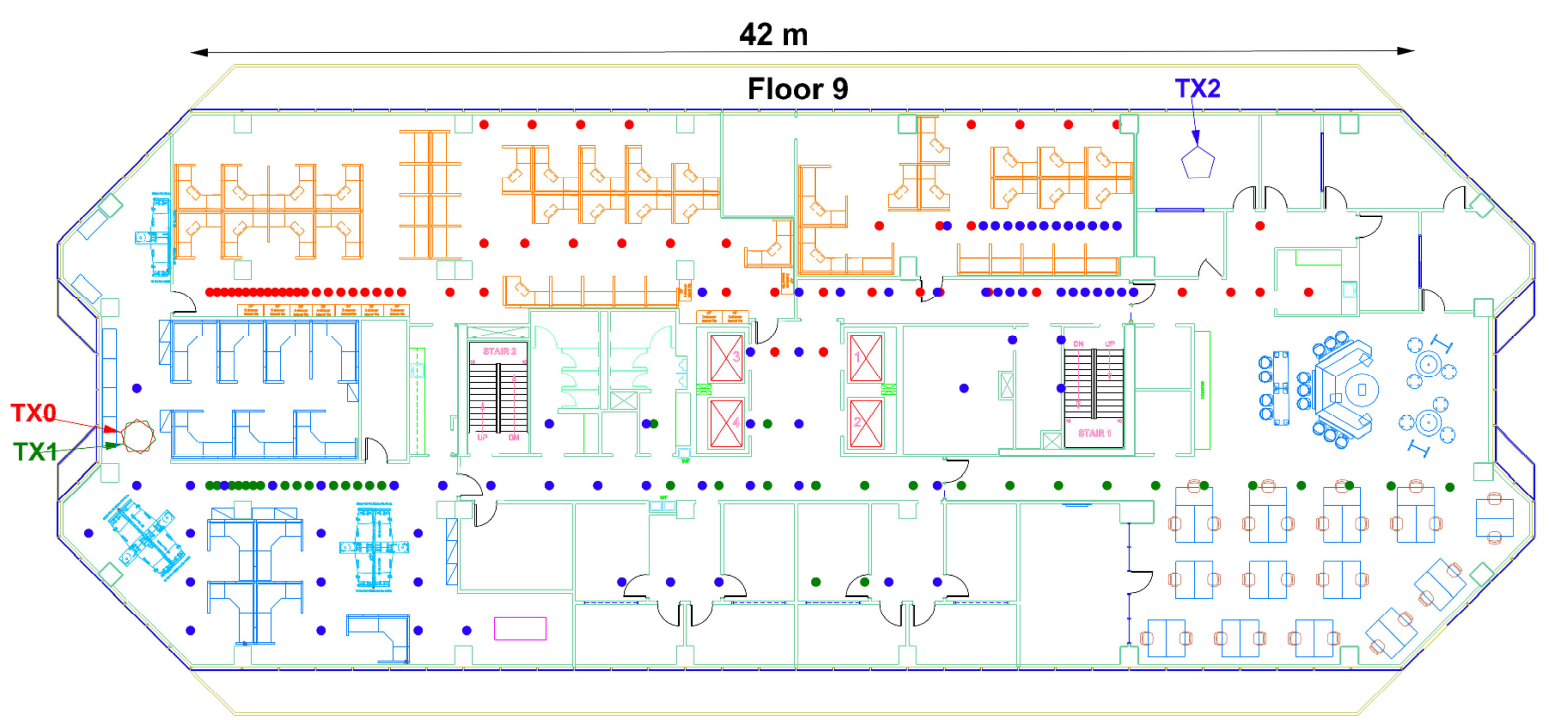} 
        \label{fig:Floor_plan}    
    }
    \caption{Showing the (a) Experiment location and office layout (b) $9^{\text{th}}$ floor office structure and transceiver positions.}
    \label{fig:bldg_arch}
    
\end{figure}

\begin{figure}[t]
\centering
  \includegraphics[clip,width=0.37\textwidth]{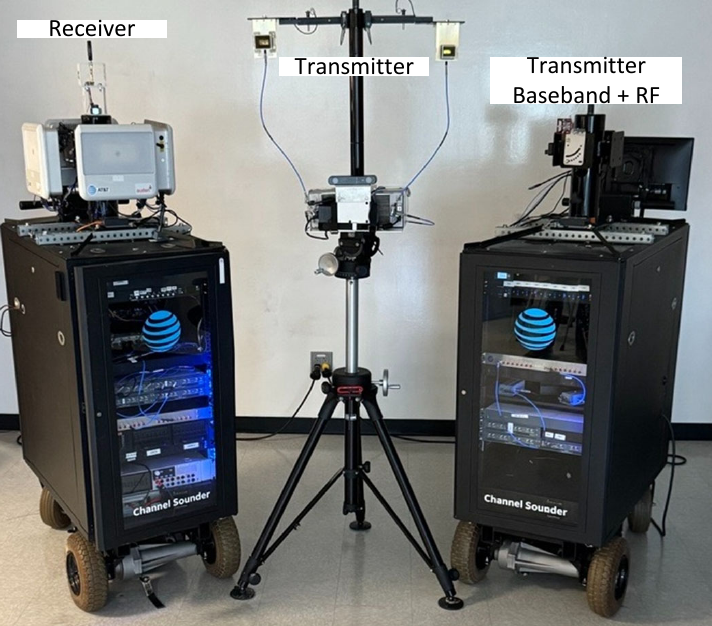}%
\caption{Indoor channel sounder mobile carts}
  \label{fig:CS-Indoor}
\end{figure}

As seen in Fig. \ref{fig:CS-Indoor}, atop the RX cart, the RF signal was concurrently received by a set of four phased array antennas and an omnidirectional antenna, placed close together to ensure that both antenna systems experienced similar wireless channel characteristics. The RF signal was downconverted to an IF frequency which was input to an SDR. The SDR demodulated the signal using a standard OFDM RX. The baseband samples were then correlated with a replica of the transmit ZC sequence to yield the complex impulse response (CIR) of the wireless channel. 

Note that at 6.9 GHz, due to the unavailability of a phased array, measurements were performed only with an omnidirectional antenna. Hence the measurements at 6.9 GHz provided valuable insights on path loss and delay spread, but could not used to compute angular spread.

The transmit (TX) and RX both had a Rubidium oscillator that generated a 10 MHz reference signal, which were synchronized via the precision time protocol (PTP) over fiber. Over-the-air calibration was performed at the beginning and end of each measurement day to ensure accuracy and repeatability of measurements \cite{Nieman_2025_01}. The specifications of the TX and RX are summarized in Tables \ref{tab:transmitter} and \ref{tab:receiver}, respectively.

\begin{table}[t!]
	\caption{TX Specification}
	\begin{center}
		\begin{tabular}{| l | c |}
			\hline
			\textbf{Attributes} & \textbf{Values}\\
			\hline \cline{1-2}
			Frequency (GHz) & 6.9, 8.3, 14.5 \\
						\hline
			Bandwidth (MHz) & 400 \\
									\hline
			Modulation & OFDM \\
									\hline
			Subcarrier spacing (kHz) & 120 \\
									\hline
			ZC Sequence Length & 3343\\
									\hline
			Antenna, Gain (dBi) & Standard horn, 10  \\
									\hline
			Antenna 3-dB beamwidth & AZ: 55$^{\circ}$, EL: 55$^{\circ}$ \\
									\hline
			Polarization & Vertical\\
									\hline
			Maximum EIRP (dBm) & +43\\
									\hline
			
			\hline
		\end{tabular}
		\label{tab:transmitter}
	\end{center}
\end{table}
\begin{table}[t!]
		\caption{RX Specification}
		\begin{center}
			\begin{tabular}{| l | c |c |}
				\hline
				\textbf{Attributes} & \multicolumn{2}{|c|}{\textbf{Values}} \\
				\hline \cline{1-3}
				Sounder & Omnidrectional & Phased Array \\
								\hline
		Center Frequency (GHz) & 6.9, 8.3, 14.5 & 8.3, 14.5 \\
								\hline
				Delay Resolution (ns) & \multicolumn{2}{|c|}{2.5} \\
								\hline
				Number of antennas & (2) Omni & (4) 8$\times$8 arrays\\
								\hline
		      	Polarization  & \multicolumn{2}{|c|}{Vertical} \\
								\hline                                
				Number of beams & N/A & 15, 30 \\
								\hline
                Field of view &  \multicolumn{2}{|c|}{AZ: {360\textdegree}, EL:{$\pm$32.5\textdegree}} \\
                			\hline
				RX type & \multicolumn{2}{|c|}{Full correlator}\\
								\hline
                    Processing Gain (dB) & \multicolumn{2}{|c|}{41}\\
								\hline
				Number of receivers & 1 & 4 (1 per array) \\
								\hline
				360\textdegree\  CIR acquisition time & $<$ 40 $\mu$s&0.5-0.9 ms\\
								\hline
				Max. Meas. Excess Delay& \multicolumn{2}{|c|}{8 $\mu$s} \\   		
						\hline
				RX noise figure& 1.5 dB & 8.3 dB \\
				\hline
			\end{tabular}
            
    			\label{tab:receiver}
		\end{center}
 
	\end{table}
\subsection{Measurement Locations and Procedures}
\label{sec:experiment}

Experiments were performed across four floors inside a commercial building located in Austin, Texas, USA shown in Fig.~\ref{fig:Arbor_bldg}. The indoor environment included office cubicles, conference rooms, hallways, and laboratory areas that were constructed with drywall with metal studs and door frames. Each floor had a similar structure with differing floor plans, room dimensions, and clutter. Fig.~\ref{fig:Floor_plan} shows the measurement locations on the $9^{th}$ floor of this building. Here, three TX locations were used (TX0, TX1, and TX2) with corresponding RX locations of each shown in matching color. 

In this work, we used the flexible multi-frequency channel sounder as described in~\cite{Nieman_2025_01} and configured for indoor use. The TX was housed in a mobile cart which was transported to a total of 11 fixed locations across four floors of a commercial building. For each TX location, the RX mobile cart traversed the floor, stopping at multiple pre-determined locations to measure the channel. Locations were chosen in LOS and NLOS conditions. The TX and RX antennas were mounted at a height of 3 m and 1.8 m, respectively. The TX antenna location was optimized for best signal coverage and to allow for measurements with the maximum possible TX-RX (TR) separation distance $d$. The experiments were conducted at a total of 653 unique locations. The RX locations were pre-selected such that T-R separations were logarithmically spaced, i.e. the close-in RX points were closer together than far-out points. To properly compere the variation across the three frequency bands, the same locations were reused for measurements across all frequency bands. All measurements in this study were conducted in accordance with the permissions granted by the FCC experimental license~\cite{fcc_license}.

\subsection{Data Processing}
\label{sec:data_processing}

The omnidirectional path loss of the indoor wireless channel is computed based on the received signal power obtained from the omnidirectional antenna. For this analysis, the power delay profile was thresholded by discarding channel taps with power levels less than 15 dB above the mean noise floor. The omnidirectional path loss at 8.3 and 14.5 GHz was also synthesized by adding the received powers obtained by the half-power-beamwidth separated, non-overlapping beams of the four phase arrays, as described in \cite{Sun_2015}. We observed that the path loss measured by both antenna systems was similar. Therefore, we use the omnidirectional antenna system to model the InH path loss, enabling us to also model the path loss at all measured frequencies.

The omnidirectional path loss ($PL$) is modeled as:
\begin{equation} \label{PL-Eq}
  PL(d) = PL_0 + 10 PLE\; \log_{10} \left( \frac{d}{d_0} \right) + S \:[dB],
\end{equation}
where $d$ is the TR separation distance in m, $PL_{\text{0}}$ is the path loss intercept at $d_{\text{0}}$ = 1 m, \textit{PLE} is the path loss exponent of the channel and $S$ is the shadow fading component and generally characterized as a zero-mean random variable with standard deviation of $\sigma_S$. 

The RMS delay spread was then computed using the data collected by the omnidirectional antenna at 6.9 GHz, and by using the data collected by the phased arrays at 8.3 and 14.5 GHz. The mean delay ($\tau_{m}$) and RMS delay spread ($\tau_{rms}$) are given by:

\begin{equation} \label{mean_dly}
  \tau_{m} = \sqrt{\frac{\sum_{i=1}^{N}\tau_{i}P_i}{{\sum_{i=1}^{N}P_i}}}, \nonumber
\end{equation}
\begin{equation} \label{rms_dly}
   \tau_{rms} = \sqrt{\frac{\sum_{i=1}^{N}(\tau_{i}-\tau_{m}^{2})P_i}{\sum_{i=1}^{N}P_i}},\nonumber
\end{equation}
where N is the number of channel taps, $\tau_i$ is the delay and $P_i$ is the power of the $i^{th}$ channel tap \cite{Rappaport_2002}.

Coherence bandwidth $B_c$ is a measure of the frequency range over which the response of the channel is approximately frequency flat. This parameter is crucial in determining the extent to which a signal will experience frequency-selective fading and is used to inform density of reference signals. It is directly calculated from the RMS delay spread as follows:

\begin{equation} \label{Cohere_BW}
 B_{c, \rho} \approx \frac{1}{K \tau_{rms}},\nonumber
\end{equation}
where $K$ is approximated as 5 or 50, if coherence bandwidth is selected such that the frequency correlation, $\rho$ is above 0.5 or 0.9, respectively \cite{Rappaport_2002}.

Angular spread of arrival ($AS$) is a measure of the spread of the power of the channel over angle and is calculated as: 
\begin{equation} \label{eq:as}
   AS = \sqrt{-2\ln\left(\left|\frac{\sum^{N}_{i=1}{e^{j\phi_{i}}}P_{i}}{\sum^{N}_{i=1}{P_{i}}}\right)\right|},\nonumber
\end{equation}
where N is the number of channel taps, and $\phi_i$ is the angle of arrival of channel tap $i$. $AS$ is defined at both the TX and the RX as the departure and arrival $AS$, respectively. In this work, using data collected by the phased arrays at 8.3, and 14.5 GHz, we computed the azimuth angle spread of arrival ($ASA$) and zenith angle spread of arrival ($ZSA$) as the spread of MPC power observed at the RX in azimuth and zenith dimensions, respectively. 

 To accurately describe certain parameters, we employ logarithmic transformations. This method is especially advantageous for data that varies across multiple scales, as it helps with compressing the spread of data. Overall, logarithmic transformations provide a powerful tool for improving the accuracy and clarity of statistical insights as is assumed in \cite{3gpp.38.901}.

\section{Key Findings}
\label{sec:KeyFindings}

In this section, we present our key findings on large-scale fading, RMS delay spread/coherence bandwidth, and angular spread. We also analyze the inter-dependencies among these parameters and their variation across frequency bands through cross-correlation analysis. Due to space limitations, Figs.~\ref{fig:SF}–\ref{fig:ZSA} focus on results at 14.5 GHz; nevertheless, these observations consistently reflect trends seen across all measured frequencies and the results are summarized in table \ref{tab:InH_Params_params} and \ref{tab:InH_Params_xcorr}.

\subsection{Large-Scale Fading}
\label{sec:fading}
Large-scale fading was analyzed by applying linear regression to fit the measured omnidirectional path loss data to Eq. (\ref{PL-Eq}). Over the four floors, for each frequency, in LOS environments, the variations in $PL_{\text{0}}$, $PLE$ and $\sigma_s$ values were limited to $\pm{0.8}\text{ dB}$, $\pm{0.14}$ and $\pm{1.3}\text{ dB}$, respectively. In NLOS environments, the variations were found to be $\pm{6}\text{ dB}$, $\pm{0.3}$ and $\pm{1.3}\text{ dB}$, respectively. Thus, at each frequency, data from all floors was combined into a single dataset to compute a large-scale fading model. 

\begin{figure}[t]
  \centering
  \subfloat[LOS]{
    \includegraphics[width=0.8\columnwidth]{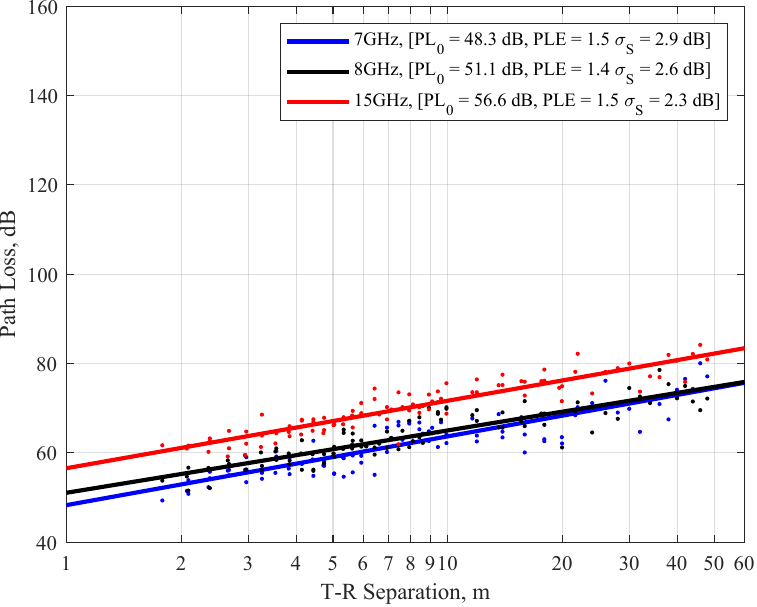} 
    
  } \\
  \subfloat[NLOS]{
    \includegraphics[width=0.8\columnwidth]{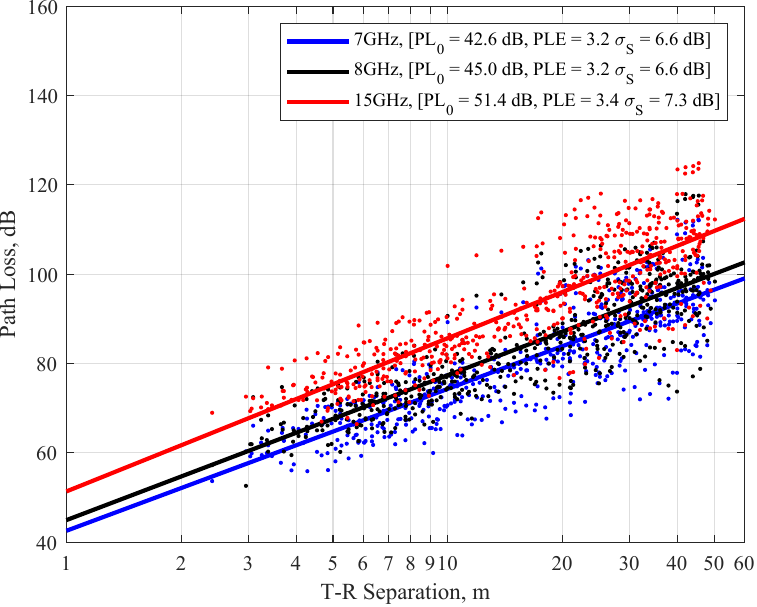} 
    
  }
  \caption{Path loss vs T-R separation for 6.9, 8.3, 14.5 GHz.}
  \label{fig:PL_vs_d}
\end{figure}

Fig.~\ref{fig:PL_vs_d} shows the scatter plot of path loss vs T-R separation ($d$) across the three measured frequency bands. All measured path loss points were within the maximum measurable path loss of the channel sounding system. The value of each estimated parameter is listed in Fig.~\ref{fig:PL_vs_d} as well as in Table~\ref{tab:InH_Params}. 

 In LOS environments, $PL_0$ was observed to be within 1 dB of free space path loss (FSPL). In NLOS environments, $PL_0$ was found to be 5-6 dB lower than that of LOS. Consequently, for close-in locations, the NLOS path loss would have been lower than the LOS path loss. Specific underlying features of the environment cause a lowering effect in $PL_0$ for NLOS environments in such a way that the PL model exhibits a higher $PLE$ with lower $PL_0$. In many scenarios observed during our experiments, the transmitted signal was not obstructed until 1-3 m beyond the TX antenna. As a result, free space propagation occurs up to the first blockage, followed by a change in the $PLE$. To correct for this breakpoint, and to ensure that the modeled NLOS path loss is at least equal to the LOS path loss, the NLOS path loss was taken to be $\max(PL_{LOS},PL_{NLOS})$ to form a two slope model \cite{3gpp.38.901}. 
 
\begin{figure}
  \centering
  \subfloat[$S$ vs. $d$]{
 		\includegraphics[width=0.8\columnwidth]{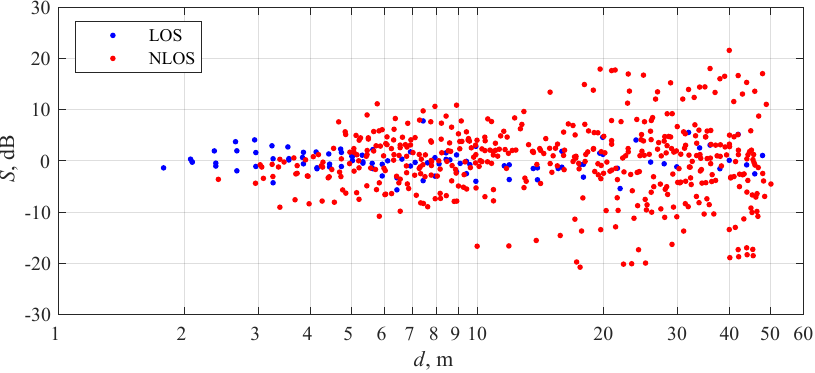}
		\label{fig:SF_vs_d}
    } \\ 
  \subfloat[Normal probability plot of $S$]{
 		\includegraphics[width=0.8\columnwidth]{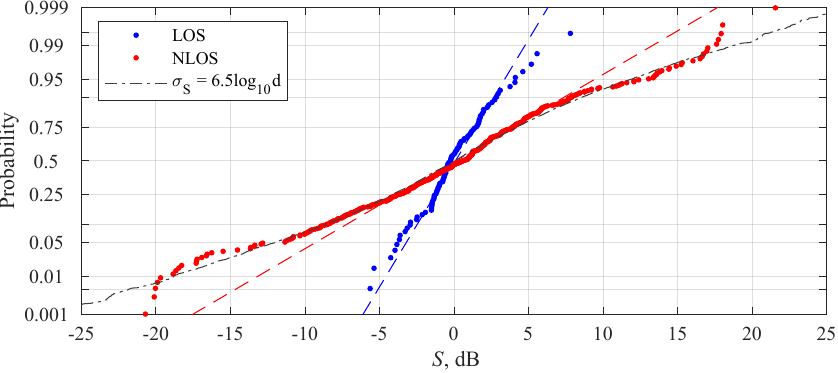}
		\label{fig:prob_plot_SF}
    }
  \caption{Shadow fading (a) vs distance (b) probability distribution for 14.3 GHz}
  \label{fig:SF}
\end{figure}

 Our measured values for $PL_0$, $PLE$, and $\sigma_S$ were similar to those provided in \cite{3gpp.38.901}. Furthermore, we observe no significant frequency dependency for the $PLE$ and $\sigma_S$. However, we observe the frequency dependency of $PL_0$ as shown in Fig. \ref{fig:PL_vs_d} and in Table \ref{tab:InH_Params}.
Shadow fading, $S$, is typically modeled as a distance independent zero-mean log-normal random variable, $S\sim\mathcal{N}(0,\sigma_S)$. 
Plots of computed $S$ vs. distance and the normal probability plot are shown in Fig. \ref{fig:SF}. For LOS, we see a close fit to normal distribution with no clear dependency on $d$. For NLOS, there appears to be a stronger dependency of $\sigma_S$ on $d$. To illustrate this dependency, we fitted a distance-dependent model of $S\sim\mathcal{N}(0,\sigma_S(d))$, where $\sigma_S(d) = 6.5\log_{10} d$. This distribution matches our data more closely as seen in Fig.~\ref{fig:prob_plot_SF}.

\label{sec:PL}

\subsection{RMS Delay Spread and Coherence Bandwidth}
The mean ($\mu$) and standard deviation ($\sigma$) of $\log_{10}(\tau_{rms}/1s)$ were found to be frequency independent. The $\mu$ values were found to be -7.92, -7.88, -7.94 within LOS regions and -7.60, -7.58, -7.59 for NLOS regions at 6.9, 8.3, and 14.5 GHz, respectively. The $\sigma$ values of $\log_{10}(\tau_{rms}/1s)$ were found to be 0.34 across all frequencies within LOS regions and 0.23, 0.21, 0.22 within NLOS regions at 6.9, 8.3, and 14.5 GHz, respectively.  Note the increase in $\mu$ values for NLOS than LOS environment. This is mainly due to increase in path lengths, multiple scattering and diffraction giving rise to a multipath rich channel. 

\begin{figure}
  \centering
  \subfloat[$\log_{10}(\tau_{rms}/1s)$ vs. $d$]{
 		\includegraphics[width=0.96\columnwidth]{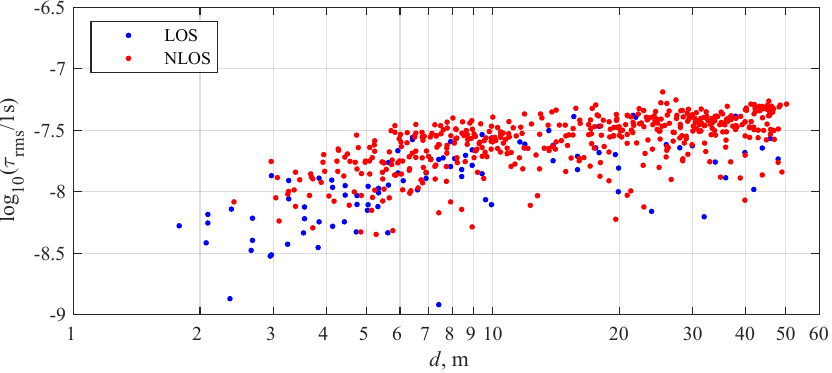}
		\label{fig:DS_vs_d}
    } \\ 
    \subfloat[Normal probability plot of $\log_{10}(\tau_{rms}/1s)$]{
  		\includegraphics[width=0.96\columnwidth]{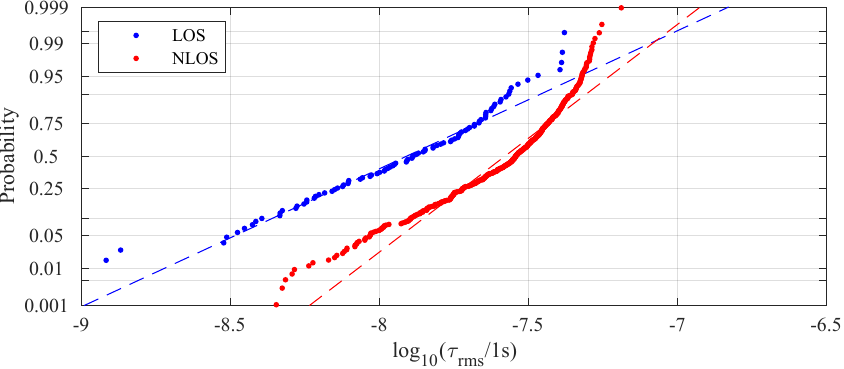}
		\label{fig:prob_plot_DS}
        }
    \caption{RMS delay spread at 14.5 GHz}
    \label{fig:DS}
\end{figure}

The difference in $\sigma$ values for LOS and NLOS environments can be best understood by examining the probability plot in Fig.~\ref{fig:prob_plot_DS}. This plot assumes log-normal distribution and shows good fit for LOS and NLOS at lower values of $\tau_{rms}$, however, there is a deviation in the upper tail of the distribution which is more pronounced for the NLOS fit. This is due to geometry of the building. The size of each floor (roughly 40 $\times$ 20 m as seen in Fig~\ref{fig:Floor_plan} limits the relative delay of MPCs with appreciable power (single reflection) that are not obstructed by structures within the building (e.g. stairwells, elevator shaft, etc.). The deviation in the upper tail of the probability narrows the distribution for NLOS leads to a smaller $\sigma$, whereas LOS and NLOS track each other more closely in the lower three quantiles.

The 90\% coherence bandwidth was found to vary from 1.5-1.7 MHz for LOS and 0.7-0.8 MHz for NLOS for measured frequency bands. The full set of delay and coherence bandwidth parameters are provided in Table~\ref{tab:InH_Params_params}.

\begin{table*}[t]
\caption{InH-Office Channel Parameters}
\label{tab:InH_Params}
\captionsetup[subtable]{justification=centering}
\hspace{0.05in}
\begin{subtable}{0.6\textwidth}
\caption{Frequency-specific parameters}
\label{tab:InH_Params_params}
\centering

\begin{tabular}{|cc|cc|cc|cc|}

\hline
\multicolumn{2}{|c|}{\multirow{2}{*}{Parameters}}                                 & \multicolumn{2}{c|}{6.9 GHz}      & \multicolumn{2}{c|}{8.3 GHz}      & \multicolumn{2}{c|}{14.5 GHz}     \\ \cline{3-8} 
\multicolumn{2}{|c|}{}                                                            & \multicolumn{1}{c|}{LOS} & NLOS & \multicolumn{1}{c|}{LOS} & NLOS & \multicolumn{1}{c|}{LOS} & NLOS \\ \hline  \cline{1-8}
\multicolumn{1}{|c|}{\multirow{3}{*}{Path Loss, dB}} & $PL_0$, dB                          & \multicolumn{1}{c|}{48.3}    &42.6      & \multicolumn{1}{c|}{51.1}    &45      & \multicolumn{1}{c|}{56.6}    & 51.4     \\ \cline{2-8} 
\multicolumn{1}{|c|}{}                               & $PLE$                        & \multicolumn{1}{c|}{1.5}    & 3.2     & \multicolumn{1}{c|}{1.4}    &3.2      & \multicolumn{1}{c|}{1.5}    &  3.4    \\ \cline{2-8} 
\multicolumn{1}{|c|}{}                               & $\sigma_S$, dB & \multicolumn{1}{c|}{2.9}    &6.6      & \multicolumn{1}{c|}{2.6}    &6.6      & \multicolumn{1}{c|}{2.3}    &7.3      \\ \hline \cline{1-8}
\multicolumn{1}{|c|}{\multirow{2}{*}{log$_{10}(\tau_{rms}/1\text{s})$}}            & $\mu$                      & \multicolumn{1}{c|}{-7.9}    &-7.6      & \multicolumn{1}{c|}{-7.9}    &-7.6      & \multicolumn{1}{c|}{-7.9}    &-7.6      \\ \cline{2-8} 
\multicolumn{1}{|c|}{}                               & $\sigma$                   & \multicolumn{1}{c|}{0.34}    & 0.23     & \multicolumn{1}{c|}{0.34}    &0.21      & \multicolumn{1}{c|}{0.34}    &0.22     \\ \hline  \cline{1-8}
\multicolumn{1}{|c|}{\multirow{2}{*}{$C_{b,\rho}$, MHz}}         & $\rho =0.5$    & \multicolumn{1}{c|}{16.5}    &8.0      & \multicolumn{1}{c|}{15.0}    &7.6      & \multicolumn{1}{c|}{17.3}    &  7.8    \\ \cline{2-8} 
\multicolumn{1}{|c|}{}                               & $\rho = 0.9$               & \multicolumn{1}{c|}{1.7}    & 0.8     & \multicolumn{1}{c|}{1.5}    &0.7      & \multicolumn{1}{c|}{1.7}    & 0.7     \\ \hline \cline{1-8}
\multicolumn{1}{|c|}{\multirow{2}{*}{log$_{10}$(ASA/1\textdegree)}}           & $\mu$            & \multicolumn{1}{c|}{--}    & --     & \multicolumn{1}{c|}{1.71}    & 1.84     & \multicolumn{1}{c|}{1.55}    & 1.77     \\ \cline{2-8} 
\multicolumn{1}{|c|}{}                               & $\sigma$                   & \multicolumn{1}{c|}{--}    & --     & \multicolumn{1}{c|}{0.13}    &  0.13    & \multicolumn{1}{c|}{0.15}    &  0.15    \\ \hline\cline{1-8}
\multicolumn{1}{|c|}{\multirow{2}{*}{log$_{10}$(ZSA/1\textdegree)}}           & $\mu$            & \multicolumn{1}{c|}{--}    & --     & \multicolumn{1}{c|}{1.24}    & 1.22     & \multicolumn{1}{c|}{1.04}    & 1.03     \\ \cline{2-8} 
\multicolumn{1}{|c|}{}                               & $\sigma$                   & \multicolumn{1}{c|}{--}    & --     & \multicolumn{1}{c|}{0.03}    &  0.03    & \multicolumn{1}{c|}{0.06}    &  0.07    \\ \hline \cline{1-8}
\multicolumn{1}{|c|}{\multirow{6}{*}{Cross-Correlations}}   & ASA vs DS           & \multicolumn{1}{c|}{--}    & --     & \multicolumn{1}{c|}{0.58}    &  0.39    & \multicolumn{1}{c|}{0.80}    &   0.44   \\ \cline{2-8} 
\multicolumn{1}{|c|}{}                               & ASA vs SF                  & \multicolumn{1}{c|}{--}    & --     & \multicolumn{1}{c|}{-0.66}    &    -0.42  & \multicolumn{1}{c|}{-0.47}    &   -0.34   \\ \cline{2-8} 
\multicolumn{1}{|c|}{}                               & DS vs SF                   & \multicolumn{1}{c|}{-0.72} &-0.55   & \multicolumn{1}{c|}{-0.54}    &   -0.51   & \multicolumn{1}{c|}{-0.39}    &  -0.30    \\ \cline{2-8} 
\multicolumn{1}{|c|}{}                               & ZSA vs SF                  & \multicolumn{1}{c|}{--}    & --     & \multicolumn{1}{c|}{-0.27}    &  -0.38    & \multicolumn{1}{c|}{0.05}    & -0.32     \\ \cline{2-8} 
\multicolumn{1}{|c|}{}                               & ZSA vs DS                  & \multicolumn{1}{c|}{--}    & --     & \multicolumn{1}{c|}{0.11}    &  -0.05    & \multicolumn{1}{c|}{0.09}    &  -0.06    \\ \cline{2-8} 
\multicolumn{1}{|c|}{}                               & ZSA vs ASA                 & \multicolumn{1}{c|}{--}    & --     & \multicolumn{1}{c|}{0.24}    &  0.18    & \multicolumn{1}{c|}{0.01}    &   0.12   \\ \hline
\end{tabular}

\end{subtable}
\hspace{0.2in}
\begin{subtable}{0.35\textwidth}
\caption{Inter-frequency cross-correlations}
\label{tab:InH_Params_xcorr}
\centering
\begin{tabular}{|cc|cc|cc|}

\hline
\multicolumn{2}{|c|}{\multirow{2}{*}{Parameters}}                                 & \multicolumn{2}{c|}{8.3 GHz}      & \multicolumn{2}{c|}{14.5 GHz}     \\ \cline{3-6} 
\multicolumn{2}{|c|}{} & \multicolumn{1}{c|}{LOS} & NLOS & \multicolumn{1}{c|}{LOS} & NLOS \\ \hline \cline{1-6}
\multicolumn{1}{|c|}{\multirow{2}{*}{6.9 GHz}} & DS & \multicolumn{1}{c|}{0.37}    &0.70      & \multicolumn{1}{c|}{0.41}    &0.70\\ \cline{2-6}
\multicolumn{1}{|c|}{}& SF & \multicolumn{1}{c|}{-0.08}    &0.91      & \multicolumn{1}{c|}{0.05}    &0.86\\ \hline \cline{1-6}
\multicolumn{1}{|c|}{\multirow{2}{*}{8.3 GHz}} & DS & \multicolumn{1}{c|}{--}    &--      & \multicolumn{1}{c|}{0.43}    &0.71\\ \cline{2-6}
\multicolumn{1}{|c|}{}& SF & \multicolumn{1}{c|}{--}    &--      & \multicolumn{1}{c|}{0.00}    &0.90\\ \hline
\end{tabular}
\end{subtable}
\end{table*}

\subsection{Azimuth and Zenith Angle Spread of Arrival}
\begin{figure}[t]
    \centering
    \subfloat[$\log_{10}(ASA/1\text{\textdegree})$ vs. $d$]{
  		\includegraphics[width=0.97\columnwidth]{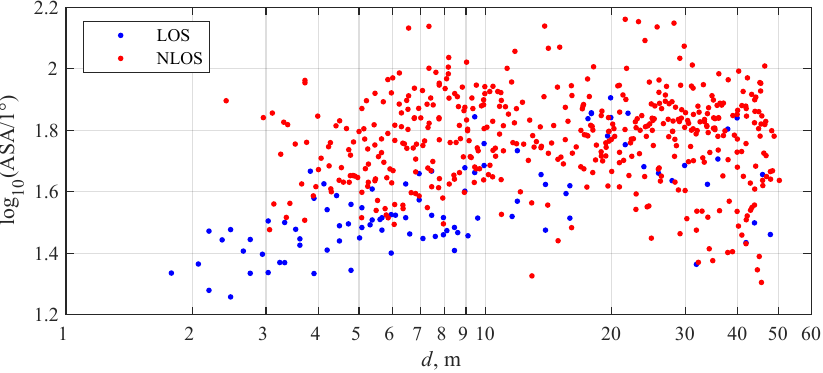}
		\label{fig:ASA_vs_d}
        } \\  
    \subfloat[Normal probability plot of $\log_{10}(ASA/1\text{\textdegree})$]{
  		\includegraphics[width=0.97\columnwidth]{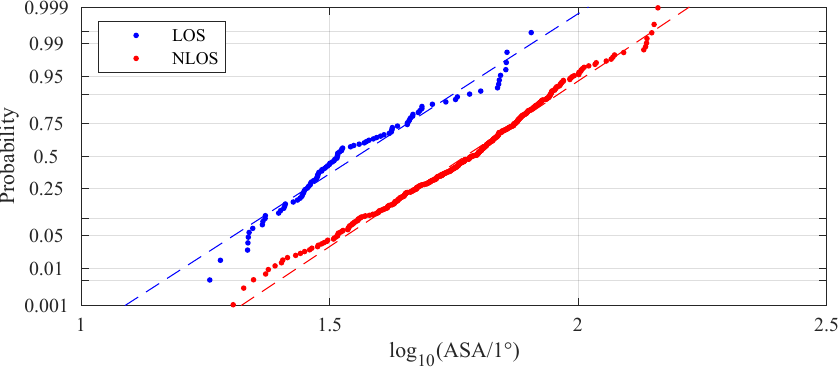}
		\label{fig:prob_plot_ASA}
        }
    \caption{ASA for 14.5 GHz}
    \label{fig:ASA}
\end{figure}

Figs.~\ref{fig:ASA} shows the scatter plot of $\log_{10}(ASA/1^\circ)$ as a function of $d$ and its normal probability plot. In particular,  in Fig.~\ref{fig:ASA_vs_d}, we note a slight increase in distance dependency of $\log_{10}(ASA/1^\circ)$. Despite this dependency, Fig.~\ref{fig:prob_plot_ASA} shows $\log_{10}(ASA/1^\circ)$ fits a log-normal distribution well for both LOS and NLOS channel conditions. The mean of $\log_{10}(ASA/1^{\circ})$ was measured as 1.71 and 1.55 for LOS and 1.84 and 1.77 for NLOS at 8.3 and 14.5 GHz, respectively. $\log_{10}(ASA/1^\circ)$ is slightly higher in NLOS vs. LOS by a factor of 1.3 for 8 GHz and 1.6 for 14.5 GHz. The standard deviation of $\log_{10}(ASA/1^\circ)$ is approximately equal for LOS and NLOS. It is 0.13, 0.15 for 8.3 and 14.5 GHz, respectively.

\begin{figure}[t]
    \centering
    \subfloat[$\log_{10}(ZSA/1\text{\textdegree})$ vs. $d$]{
  		\includegraphics[width=0.97\columnwidth]{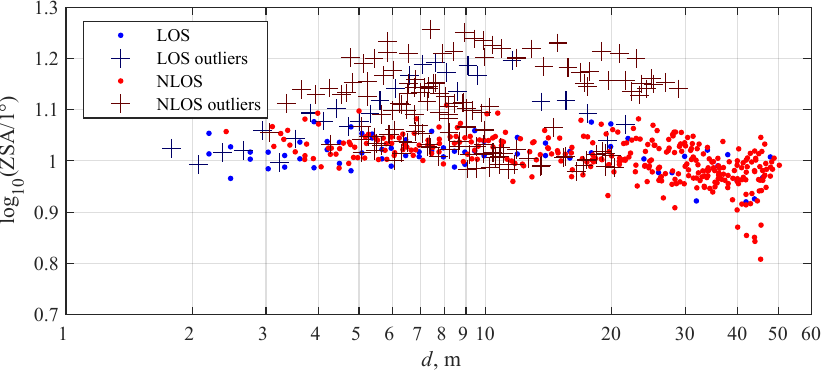}
		\label{fig:ZSA_vs_d}
        } \\  
    \subfloat[Normal probability plot of $\log_{10}(ZSA/1\text{\textdegree})$]{
  		\includegraphics[width=0.97\columnwidth]{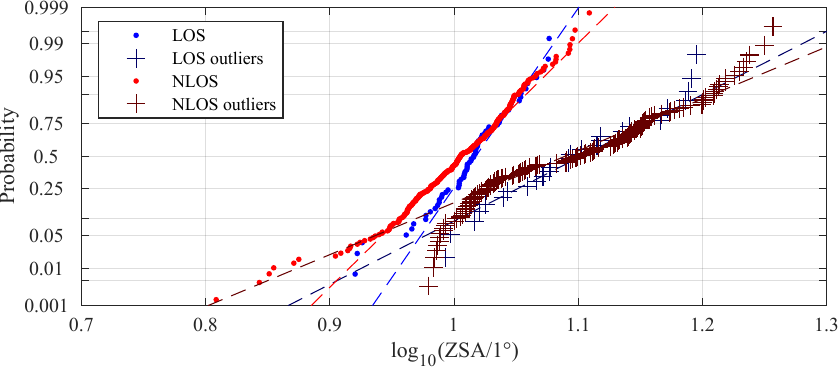}
		\label{fig:prob_plot_ZSA}
        }

    \caption{ZSA for 14.5 GHz}
        \label{fig:ZSA}
\end{figure}

Figs. \ref{fig:ZSA_vs_d} and \ref{fig:prob_plot_ZSA} show $\log_{10}(ZSA/1^\circ)$ as a function of $d$ and its probability plot, respectively. We found that ZSA over the full dataset was an ill-fit to a single log-normal distribution. In our data, ZSA is a much better fit to a two-mode distribution with 2/3 of the data fitting the first mode ($\mu \approx 10\text{\textdegree}$) and the remaining 1/3 following the second mode ($\mu \approx 13\text{\textdegree}$). The two modes are designated via the lighter (first mode) and darker (second mode) points. Upon further investigation, we found two major reasons for this. In LOS channel conditions, wider ZSA points occurred in lab areas with raised floors. This changed channel geometry vs. underlying building structure (relative height of receiver increased by 0.6 m with respect to concrete subfloor) leading to higher ZSA. In NLOS channel conditions, we observe higher ZSA points in locations where there is significant nearby blockage. For example, in a section of 9$^th$ floor points, there was a wall of 2 m high filing cabinets within 0.4 m of RX phased array antenna. In another section of points on the 3$^{rd}$ floor, computing room equipment coolers caused similar blockage. The  mean and standard deviation of ZSA are reported in Table~\ref{tab:InH_Params_params}.    

Regarding distance dependency, in contrast to azimuth angular spread, the ZSA values appear to have a negative dependency with $d$ (see Fig.~\ref{fig:ZSA_vs_d}). This is likely due to the physical constraints of the building (concrete and steel floors and ceilings), which begin to limit the zenith extent of spread of energy like a waveguide as $d$ increases. A similar trend was observed for 8.3 GHz data. With regard to frequency depoendency, we observe reduced angular spread in both dimensions as frequency increases. Refer to Table~\ref{tab:InH_Params_params} for full summary of angular parameters.

\subsection{Inter-Parameter Cross-Correlations}
Table~\ref{tab:InH_Params_params} shows the cross-correlations of logarithm of parameters for each measured frequency. ASA and DS are found to have positive correlation for both LOS and NLOS channel conditions. Shadow fading is negatively correlated with both ASA and DS. The degree of correlation for shadow fading appears to decrease as frequency increases. ZSA appears to be uncorrelated with other parameters.

\subsection{Inter-Frequency Cross-Correlations}
Table~\ref{tab:InH_Params_xcorr} shows the cross-correlations of logarithm of large-scale parameters over frequency. RMS delay spread was observed to have positive correlation with frequency in both LOS and NLOS environments with a much higher correlation observed for NLOS. For shadow fading, we observe approximately zero correlation in LOS with high correlation in NLOS. For LOS, we believe this is due to the constructive and destructive interference between strong MPCs present in LOS conditions, leading to variation of power that is a function of channel geometry as well as the wavelength. This variation is captured as shadow fading parameter $\sigma_S$ from~(\ref{PL-Eq}) and is uncorrelated for the frequencies measured. We suspect that for simpler channel geometries with wavelengths that are small integer multiples of each other, a correlation of $S$ with frequency may be observed. In NLOS environments, however, we observed that shadow fading is highly correlated over frequency, since primary propagation features of the NLOS channel---e.g. blockage, penetration loss, reflection coefficient, etc, are frequency agnostic.

\section{Conclusions}
\label{sec:conclusions}
In this paper, we present a comprehensive study of large scale wireless channel in indoor-office environment over three FR3 frequency bands, at 6.9, 8.3, and 14.5 GHz. For each band, we characterized path loss, shadowing, RMS delay spread, and large-scale angular characteristics. Our data set allows us to not only quantify cross-correlations between parameters but also cross-correlations of parameters vs. frequency. These findings may be used to refine indoor wireless FR3 channel models, ultimately improving network design and performance in commercial buildings.

\bibliographystyle{IEEEtran}
\bibliography{references}

\end{document}